\documentstyle[prb,aps,multicol,epsf]{revtex}
\begin{document}
\input epsf 
\title{Vortex with Fractional Quantum Numbers  
in Chiral $p$-Wave Superconductor}
\author{J. Goryo} 
\address{\it Department of Physics, Hokkaido University, 
Sapporo, 060-0810 Japan}
\date{\today}
\maketitle

\begin{abstract}

We show that a vortex in a chiral $p$-wave superconductor, which has   
the $p_{x}+ i p_{y}$-wave pairing state and breaks   
$U(1)$, parity and time reversal symmetry simultaneously,  
has fractional charge $-\frac{n e}{4}$ and 
fractional angular momentum $-\frac{n^{2}}{16}$ ($n$; vorticity). 
This suggests that the vortex could be anyon and 
could obey fractional statistics.  
Electromagnetic property of the vortex is also 
discussed and we find that an electric field is induced 
near the vortex core.

\end{abstract}
\draft

%\newpage
\begin{multicols}{2}

\section{Introduction}

Recently, the superconductivity of $Sr_{2}RuO_{4}$ was discovered 
by Maeno {\it et. al.}.\cite{Maeno}
This superconductivity seems to exhibit  
chiral $p$-wave order, in which the symmetry of the wave function of 
Cooper pair is $p_{x} + i p_{y}$-wave \cite{Rice-Sigrist} and 
parity (P) and time reversal symmetry (T) are violated. 
In this paper, we show that a vortex in chiral $p$-wave superconductor 
has fractional charge and fractional angular momentum 
as the consequence of P- and T-violation.

$Sr_{2}RuO_{4}$ has a layered 
perovskite crystal structure without cupper and has low transition temperature 
$T_{c} \simeq 1.5[K]$. 
$Sr_{2}RuO_{4}$ is a strongly correlated 2-dimensional Fermi liquid
\cite{Mackenzie,Inoue,Maeno-2} and is partially similar to 
$^{3}$He which shows p-wave superfluidity.\cite{He-A text} 
The evidence for unconventional ({\it i.e.}, non s-wave) superconductivity 
was shown by the absence of a Habel-Slichter peak in $1/T_{1}T$ of 
NQR-measurements\cite{NQR} and $T_{c}$-suppression 
by non-magnetic impurities.\cite{non-mag-imp} 
Rice and Sigrist proposed that pairing symmetry of orbital part 
is chiral $p$-wave,\cite{Rice-Sigrist} 
which has the same orbital symmetry of 
superfluid $^{3}$He-A.\cite{He-A text}
The proposal is consistent with the discovery of internal magnetic field 
in the superconducting phase by $\mu$SR measurement\cite{muSR} 
and the experiment of $^{17}$O-NMR Knight shift.\cite{NMR} 
In the chiral $p$-wave superconductor, Cooper pair has orbital angular 
momentum $l_{z}=1$, where $z$ is perpendicular to 
the superconducting plane, {\it i.e.}, the same direction as   
the c axis of the crystal. Therefore, local $U(1)$ gauge symmetry, 
P and T are spontaneously broken 
by the same order parameter.

The above situation is analogous to the quantum Hall system (QHS), 
which is the 2-dimensional electron system in 
an external magnetic field.\cite{Klitzing}  
In QHS the external magnetic field violates P and T,    
but the explicit local $U(1)$ gauge symmetry is preserved.  
It was shown that the Chern-Simons term is 
induced in the effective action of the gauge field 
when Fermion is integrated out.\cite{2+1C-S,Ishikawa} 
The Chern-Simons term is a P- and T-odd bi-linear form of 
the gauge fields and has one derivative. Therefore, this term 
plays important roles in low energy and long distance physics 
and causes P- and T-violating electromagnetic phenomena.  
The coefficient of the induced Chern-Simons term is 
quantized exactly as (integer)$\times$ $\frac{e^{2}}{2 \pi}$. 
$U(1)$ gauge invariance guarantees the exact quantization. 
Since the coefficient of the term becomes 
the Hall conductance, the integer quantum Hall effect was  
explained from gauge invariance.\cite{Ishikawa,QHE1,QHE2}
     
It was shown that the Chern-Simons term is also
induced in the Ginzburg-Landau action of chiral $p$-wave 
superconductors.\cite{Goryo-Ishikawa,Volovik-G-I} In this case, 
the Chern-Simons term does not have exactly quantized coefficient 
because of the spontaneous 
breaking of local $U(1)$ gauge symmetry. The coefficient is 
approximately equal to the fine structure constant.\cite{Goryo-Ishikawa} 

In this paper, we discuss a vortex in chiral $p$-wave 
superconductor starting from the Ginzburg-Landau Lagrangian\cite{Matsu-Sig}  
with the Chern-Simons term. Such a vortex could obey 
unusual statistics by the existence of the Chern-Simons term. 
The above possibility has been pointed out in Chern-Simons-Higgs model 
by several authors.\cite{A-H-C-S} In this model, 
vortex solutions have been found and they have  
fractional angular momentum and fractional charge.  
The Chern-Simons vortex is a candidate of anyon that obeys 
the fractional statistics due to 
fractional angular momentum.\cite{Wilczek}   
In the present paper we find 
a vortex solution in chiral $p$-wave superconductor 
and show that it has fractional charge $-\frac{n e}{4}$ and 
fractional angular momentum $-\frac{n^{2}}{16}$ with 
integrally quantized magnetic flux $\frac{2 n \pi}{2 e}$ 
($n=0,\pm 1,\pm 2 \cdot\cdot\cdot$; vorticity)\cite{V-S-L},
hence the vortex could be anyon.     
The starting Ginzburg-Landau Lagrangian with 
the Chern-Simons term is 
an extended version of the Chern-Simons-Higgs model, so to say, 
the non-relativistic Maxwell-Chern-Simons-vector Higgs model.
We also discuss electromagnetic property of the vortex. 
In the region $r < \lambda$ ($r$ is the distance from the vortex core, 
and $\lambda$ is London penetration depth), the vortex has  
an electric field from the Chern-Simons term 
and it would be expected that non-trivial electromagnetic 
phenomena occur. $Sr_{2}RuO_{4}$ is known as a type II 
superconductor,\cite{Yoshida} therefore it has a stable vortex. 
It is interesting if these exotic properties of the vortex   
are observed experimentally in the superconductivity of $Sr_{2}RuO_{4}$.  

The paper is organized in the following manner. 
In section II, we review the derivation of the 
Ginzburg-Landau Lagrangian with the Chern-Simons term 
in chiral $p$-wave superconductor. 
In section III, we study charge and angular momentum of a vortex. 
In section IV, we discuss electromagnetic property of the vortex
and find that an electric field is induced near the vortex core. 
Summary is given in section V. We use the natural unit ($\hbar=c=1$) 
in the present paper.    

\section{Derivation of the Chern-Simons term and 
the Ginzburg-Landau Lagrangian of chiral $p$-wave superconductor} 

Following Ref.\cite{Goryo-Ishikawa}, we review how the Chern-Simons term 
is induced in the Ginzburg-Landau Lagrangian of  
chiral $p$-wave superconductor 
such as $Sr_{2}RuO_{4}$ in this section.
At first, we consider one superconducting layer, 
and we extend our discussion in multi-layer system. 

Fermionic Lagrangian of superconductor is written as 
\begin{eqnarray}
{\cal {L}}&=&-\frac{1}{4}F^{\mu\nu}F_{\mu\nu} + 
\Psi^{\dagger} (i \partial_{0} - e A_{0} \tau_{3}) \Psi
\nonumber\\ 
&&- \Psi^{\dagger} 
\left( \begin{array}{cc} 
\epsilon({\bf p}+e{\bf A}) & \Delta ({\bf p}) \\
\Delta^{\dagger} ({\bf p}) & - \epsilon({\bf p} - e {\bf A}) 
\end{array} \right)  \Psi , 
\label{N-B-L}
\end{eqnarray}
where 
$\Psi_{\alpha}=\frac{1}{\sqrt{2}}\left(\begin{array}{c} \psi_{s}\\ 
\psi^{\dagger}_{s} 
\end{array} \right)$ 
is Fermion field in Nambu-Bogoliubov representation\cite{Nambu} 
with ``isospin'' $\alpha$, and $\psi_{s}$ is 
the usual representation for Fermion field with spin index $s$. 
$({\vec {\tau}})_{\alpha\beta}$ is Pauli matrix 
with isospin indices, and $\epsilon ({\bf p})$ is the kinetic energy of 
Fermion and is measured from Fermi energy $\epsilon_{\rm F}$.   
For simplicity, we choose 1 band model with cylindrical  
Fermi surface,\cite{Mackenzie} therefore, $\epsilon({\bf p})$ 
is written as 
$\epsilon({\bf p})=\frac{p^{2}}{2 m_{e}} - \epsilon_{\rm F}$ with 
electron effective mass $m_{e}$. 
$\Delta({\bf p}) \sim V_{pair}<\psi \psi>$ is the gap function 
($<\cdot\cdot\cdot>$ denotes the expectation value of operator 
in the ground state and 
$V_{pair}$ shows electron-electron interaction by which pairing state arises) 
and it is written as  
\begin{equation}
\Delta({\bf p}) = i {\vec {\sigma}} \sigma_{2} \cdot {\vec {d}}({\bf p}),  
\end{equation}
in the p-wave (spin-triplet) pairing state.\cite{Sigrist-Ueda} 
${\vec{\sigma}}_{ss^{\prime}}$ is Pauli's spin matrix and ${\vec{d}}$ is 
so-called $d$-vector. 
In the superconductivity of $Sr_{2}RuO_{4}$,  
proposed $d$-vector is written as\cite{Rice-Sigrist}  
\begin{equation}
\left\{ \begin{array}{ccl}
{\vec{d}}({\bf p})&=&\hat{{\vec{z}}} {\bf \eta}\cdot\hat{\bf p},\\
{\bf \eta({\bf x})}&=&(\eta_{x}({\bf x}),\eta_{y}({\bf x})),   
\end{array}\right. 
\label{d-vector}
\end{equation}      
where ${\bf \eta}$ is a vector order parameter field and 
it has an expectation value written as 
\begin{equation}
<{\bf \eta}>=|\eta_{0}|(1,i),   
\end{equation}   
which shows chiral $p$-wave pairing state. 
For a while, we neglect the spatial dependence of ${\bf \eta}$ and put 
${\bf \eta}=<{\bf \eta}>$. 

Gauge transformation of $\Psi$ and $A_{\mu}$ is  
written as 
\begin{equation}
\left\{ \begin{array}{ccc}
\Psi &\rightarrow& e^{- i e \tau_{3} \varphi} \Psi, \\
A_{\mu} & \rightarrow& A_{\mu} + \partial_{\mu} \varphi.    
\end{array} \right. 
\end{equation}
For convenience, we take 
$\partial_{0} A_{0} - c_{s}^{2} \partial_{i} A_{i}=0$ gauge, 
where $c_{s}$ is the velocity of $U(1)$ Goldstone mode, 
and fix the redundant gauge degree of freedom which 
corresponds to the degree of freedom of $U(1)$ Goldstone mode.  

It is known that the massive particle 
gives no higher order corrections to 
the Chern-Simons term, but the massless 
particle may give a correction.\cite{Coleman-Hill}   
To find out a higher order correction, the authors of 
Ref.\cite{Goryo-Ishikawa} considered a correction to the induced Chern-Simons 
term by $U(1)$ Goldstone mode. 
It was shown that $U(1)$ Goldstone mode does not give a correction 
to the induced Chern-Simons term and its degree of freedom corresponds 
to the redundant gauge degree of freedom of the system.

%it guarantees the gauge invariance of the 
%lagrangian, as usuall\cite{Nambu}. 

From Eq.(\ref{N-B-L}), we can obtain Fermion propagator and calculate 
Fermion loop. Before doing that, to simplify notations, 
we introduce a vector ${\vec{g}}({\bf p})$ as 
\begin{equation}
{\vec{g}}({\bf {p}})= 
\left(\begin{array}{c} 
{\rm Re}\Delta({\bf {p}}) \\ 
-{\rm Im}\Delta({\bf {p}}) \\
\epsilon ({\bf {p}}) 
\end{array} \right)= 
\left(\begin{array}{c} 
|\eta_{0}|\hat{p_{x}} i \sigma_{3} \sigma_{2} \\ 
-|\eta_{0}|\hat{p_{y}} i \sigma_{3} \sigma_{2} \\
\epsilon ({\bf {p}}) 
\end{array} \right). 
\label{g-vec}
\end{equation} 
By using ${\vec{g}}({\bf p})$, Eq. (\ref{N-B-L}) is rewritten as  
\begin{equation}
\left\{\begin{array}{ccl}
{\cal{L}}&=&\Psi^{\dagger} \left[i \partial_{0} - 
{\vec{\tau}}\cdot{\vec{g}}({\bf p}) \right] \Psi - j_{\mu} A^{\mu}, \\
j_{0}({\bf q})&=&
e \sum \Psi_{\bf p}^{\dagger} \tau_{3} \Psi_{{\bf p}+{\bf q}},  \\
{\bf j}({\bf q}) &=&{\bf j}^{(p)}({\bf q}) + {\bf j}^{(d)}({\bf q}), 
\end{array}\right. 
\label{re-N-B-L}
\end{equation}
where ${\bf j}^{(p)}$ is the paramagnetic current and ${\bf j}^{(d)}$ is 
the diamagnetic current. They are written as 
\begin{equation}
\left\{ \begin{array}{ccl}
{\bf j}^{(p)}({\bf q})&=&
e \sum \Psi_{\bf p}^{\dagger} 
\{ \frac{\partial}{\partial \bf p} g_{3}({\bf p} + \frac{\bf q}{2}) \} 
\Psi_{{\bf p} + {\bf q}},
\\  
{\bf j}^{(d)}({\bf q})&=&
\frac{e^{2}}{m_{e}} \sum \Psi_{\bf p}^{\dagger} {\bf A}({\bf q}) \tau_{3} 
\Psi_{\bf p+q}.     
\end{array}\right. 
\end{equation}
We regard the interaction $j_{\mu}A^{\mu}$ as a perturbation and 
calculate correlation functions.  
Fermion propagator 
$G_{{\bf p}}(t-t^{\prime})
=<T\Psi_{\bf p}(t)\Psi_{\bf p}^{\dagger}(t^{\prime})>$ 
($T$ means the T product) 
obtained from Lagrangian Eq. (\ref{re-N-B-L}) is written 
after the Fourier transformation with (2+1)-dimensional spacetime 
momentum $p_{\mu}=(p_{0},{\bf p})$ as 
\begin{eqnarray}
G(p)&=&(p_{0} - {\vec {\tau}} \cdot {\vec g}({\bf p}))^{-1} \nonumber\\
&=&\frac{p_{0} + {\vec {\tau}} \cdot {\vec g}({\bf p})}{p_{0}^{2} - 
|\vec{g}|^{2} + i \delta},   
\end{eqnarray}
where $i \delta$ stands for a infinitesimal imaginary constant 
and it fixes the boundary condition 
for the propagator.   
  
The Fourier transforms of 
density-density, density-current, and current-current correlation 
functions in the lowest order of the perturbative expansion are written as 
\begin{eqnarray}
\pi_{00}(q)&=&-i(F.T.)<Tj_{0}(x)j_{0}(x^{\prime})> 
\nonumber\\
&=& 
- i e^{2} \int 
\frac{d^{3}p}{(2 \pi)^{3}} Tr\left[\tau_{3} G(p+q) \tau_{3} G(p) \right]
\nonumber\\ 
&=&e^{2} m_{e} + {\cal{O}}(q^{2}) , 
\label{pi00}\\
\pi_{0i}(q)&=&-i(F.T.)<Tj_{0}(x)j_{i}^{(p)}(x^{\prime})>
\nonumber\\   
&=&-i e^{2} \int 
\frac{d^{3}p}{(2 \pi)^{3}} Tr\left[\tau_{3} G(p+q) 
(- \frac{g_{3}({\bf p})}{\partial p^{i}}) G(p) \right] 
\nonumber\\
&=&- i \sigma_{xy} \varepsilon_{ij} q^{j} + {\cal{O}}(q^{2}), 
\label{pi0i}\\
\pi_{ij}(q)&=&-i(F.T.)<Tj_{i}^{(p)}(x)j_{j}^{(p)}(x^{\prime})> 
\nonumber\\
&=&-i e^{2} \int 
\frac{d^{3}p}{(2 \pi)^{3}} \frac{g_{3}({\bf p})}{\partial p^{i}}
\frac{g_{3}({\bf p})}{\partial p^{j}}Tr\left[G(p+q) G(p)\right] 
\nonumber\\
&=&{\cal{O}}(q^{2}) ,
\label{piij}
\end{eqnarray}
and a contribution from the diamagnetic current in the same order is  
\begin{eqnarray}
\pi_{ij}^{(d)}(q)
&=&(F.T.)\frac{\delta}{\delta A^{j}(x^{\prime})}<j_{i}^{(d)}(x)>
\nonumber\\
&=&-\frac{e^{2} \rho_{e}}{m_{e}}\delta_{ij},  
\label{piijd}\end{eqnarray}
where $x$ denotes the coordinate in (2+1)-dimensional spacetime, 
$(F.T.)(\cdot\cdot\cdot)$ means that  
we make the Fourier transformation of following terms,   
$Tr$ means the trace about isospin and {\it real} spin indices, and 
$\rho_{e}=\frac{m_{e} \epsilon_{F}}{\pi}$ is Fermion number density.  
Since we are interested in low energy and long distance region, we 
neglect ${\cal {O}}(q^{2})$ terms.  
The constant $\sigma_{xy}$ will be the coefficient of the Chern-Simons term,  
and it is written as  
\begin{eqnarray}   
\sigma_{xy}&=&
\frac{i}{2!} \varepsilon^{ij} 
\frac{\partial}{\partial q^{j}} \pi_{0i}(q) |_{q=0}
\nonumber\\
&=&\frac{e^{2}}{8}\int \frac{d^{2} p}{(2 \pi)^{2}} 
\frac{tr[\vec{g} \cdot ({\bf \partial}\vec{g} \times {\bf \partial}\vec{g})
-g_{3} ({\bf \partial}\vec{g} \times {\bf \partial}\vec{g})_{3}]}
{tr[\frac{1}{2}{\vec g} \cdot {\vec g}]^{\frac{3}{2}}}, \label{Hallcond3}
\end{eqnarray}
here $tr$ means trace about {\it real} spin indices and 
${\bf \partial}=\frac{\partial}{\partial {\bf p}}$. 
The first term in Eq.(\ref{Hallcond3}) is a topological 
invariant. Volovik argued this type of topological invariant. 
\cite{Volovik-G-I,Volovik} 
The second term is not a topological invariant. 
The second term appears because $U(1)$ gauge symmetry is 
spontaneously broken.\cite{Goryo-Ishikawa}

By substituting Eq. (\ref{g-vec}) into Eq. (\ref{Hallcond3}), we 
obtain the value 
\begin{equation}
\sigma_{xy}=\frac{e^{2}}{4 \pi}, 
%\frac{\epsilon}{|\epsilon|} 
%\nonumber\\
%&=&\frac{e^{2}}{4 \pi^{2}}
%\frac{-i <\eta>^{*} \times <\eta>}{|<\eta>^{*} \times <\eta>|}.   
\label{sigma_xy}
\end{equation}
which coincides with the fine structure constant. 
In this case, 
the non-topological term in Eq. (\ref{Hallcond3}) vanishes 
accidentally. In general, 
the non-topological term does not vanish. Actually, if we calculate 
$\sigma_{xy}$ in the tight binding scheme, the non-topological term 
has negligibly small value and $\sigma_{xy}$ is approximately equal 
to the fine structure constant.\cite{Goryo-Ishikawa} 

From Eqs. (\ref{pi00}), (\ref{pi0i}), (\ref{piij}), and (\ref{piijd}),  
the low energy effective Lagrangian in the case that the 
vector order parameter field is constant ({\it i.e.}, 
${\bf \eta}({\bf x})=<{\bf \eta}>=|\eta_{0}|(1,i))$ is written as 
\begin{eqnarray}   
{\cal{L}}_{eff.}&=&-\frac{1}{4}F^{\mu\nu}F_{\mu\nu} + 
\frac{1}{2} (\frac{m^{2}}{c_{s}^{2}} A_{0}^{2} - m^{2} {\bf A}^{2})
\nonumber\\
&&+\frac{\sigma_{xy}}{2}
\varepsilon_{0ij}(A_{0}\partial_{i}A_{j}+A_{i}\partial_{j}A_{0})
\nonumber\\ 
&&+ {\cal{O}}(e^{3}).  
\label{eff-lag}
\end{eqnarray}
The second term in the {\it r.h.s.} of Eq.(\ref{eff-lag}) contributes  
Thomas-Fermi screening, and the third term contributes Meissner effect. 
The parameters $m$ and $c_{s}$ are determined as  
$
m=(\frac{\rho_{e} e^{2}}{m_{\rm e}})^{1/2} ~,~ c_{s} 
\simeq (\frac{\rho_{e}}{m_{\rm e}^{2}})^{1/2}=
(\frac{{v_{\rm F}}^{2}}{2 \pi})^{1/2}, 
$ 
where $m_{\rm e}$, $\rho_{e}$ and $v_{\rm F}$ are mass, number density 
and Fermi velocity of electrons, respectively. 
$\lambda=m^{-1}$ is the penetration depth for the magnetic field in the 
usual case ({\it i.e.} $\sigma_{xy}=0$).  
The combination of the forth term and the fifth term 
in the {\it r.h.s.} of Eq.(\ref{eff-lag}) is 
the Chern-Simons term in 
$\partial_{0} A_{0} - c_{s}^{2} \partial_{i} A_{i}=0$ gauge. 

Next we take into account spatial dependence of 
order parameter fields. 
Actual superconductor is 
spatially 3-dimensional object which consists of layer of 
the superconducting plane. 
Therefore, we extend above discussion to layered system. 
In low energy and long distance 
region, it is sufficient to consider lower mass dimensional terms.   
Except for the Chern-Simons term, 
the quite general phenomenological Ginzburg-Landau free energy 
of the chiral $p$-wave superconductor with tetragonal symmetry 
which corresponds to the layered perovskite structure of the crystal lattice 
has been proposed by Sigrist and Ueda.\cite{Sigrist-Ueda}
We combine both of them. Lagrangian would have the following form ;   
\begin{eqnarray} 
{\cal {L}}_{G-L}&=&n_{p} \frac{\sigma_{xy}}{2}
\epsilon^{ij3}(A_{0} \partial_{i} A_{j} + A_{i} \partial_{j} A_{0})
-\frac{1}{4} F^{\mu\nu}F_{\mu\nu} 
\nonumber\\
&&+ |D_{0} {\bf \eta}|^{2} 
- K_{1}(|D_{x} \eta_{1}|^{2}+|D_{y} \eta_{2}|^{2})
\nonumber\\
&&- K_{2}(|D_{y} \eta_{1}|^{2}+|D_{x} \eta_{2}|^{2})
\nonumber\\
&&- K_{3}\left\{(D_{x} \eta_{1})^{*}(D_{y} \eta_{2}|)+c.c.\right\}
\nonumber\\
&&- K_{4}\left\{(D_{x} \eta_{2})^{*}(D_{y} \eta_{1}|)+c.c.\right\}
- K_{5}|D_{z} {\bf \eta}|^{2} 
\nonumber\\
&&-A|{\bf \eta}|^{2} - \beta_{1}|{\bf \eta}|^{4} 
- \beta_{2} |{\bf \eta}^{*} \times {\bf \eta}|^{2} 
\nonumber\\
&&- \beta_{3}|\eta_{1}|^{2}|\eta_{2}|^{2} + C,   
\label{G-G-L-C-S-0}
\end{eqnarray}
where $D_{\mu}=\partial_{\mu} + 2 i e A_{\mu}$, 
$n_{p}$ is a number of the superconducting 
plane per unit length along the $z$ axis (the $c$ axis), 
and $A$, $K_{i}$, and $\beta_{i}$, are phenomenological constants 
with appropriate dimensions.

To simplify the calculations, we use    
\begin{equation}
\left\{ \begin{array}{c}
K_{1}=K_{2}=c_{s}^{2},~ K_{3}=-K_{4}=c_{s}^{2} \gamma, \\
\beta_{1}=\frac{a}{4}+b, ~\beta_{2}=-b, ~\beta_{3}=0,\\     
A=-\frac{a |\eta_{0}|^{2}}{2}, ~C=\frac{a |\eta_{0}|^{4}}{4}, ~(0<a,b) . 
\end{array} \right. 
\label{parameters-Sig.-Ueda}
\end{equation}
These assignments of the phenomenological parameters in 
Eq. (\ref{parameters-Sig.-Ueda}) seems artificial\cite{Agterberg}, however, 
%as far as charge and angular momentum of a vortex are concerned, 
%our argument is quite general because 
charge and angular momentum of a vortex do not depend on 
these phenomenological parameters as will become clear in the next section. 
Eq. (\ref{G-G-L-C-S-0}) reduces its form with 
cylindrical symmetry to 
\begin{eqnarray} 
{\cal {L}}_{G-L}&=&n_{p} \frac{\sigma_{xy}}{2}
\epsilon^{ij3}(A_{0} \partial_{i} A_{j} + A_{i} \partial_{j} A_{0})
-\frac{1}{4} F^{\mu\nu}F_{\mu\nu} 
\nonumber\\
&&+ |D_{0} {\bf \eta}|^{2} 
- c_{s}^{2}|{\bf D}_{//} {\bf \eta}|^{2} 
-c_{s}^{2}\gamma\epsilon^{ij3}\epsilon^{kl}(D_{i} \eta_{k})^{*} D_{j} \eta_{l} 
\nonumber\\
&&- K_{5} |D_{z} {\bf \eta}|^{2} 
\nonumber\\
&&
-\frac{a}{4}(|{\bf \eta}|^{2} - |\eta_{0}|^{2})^{2} - 
b (|{\bf \eta}|^{4} - |{\bf \eta}^{*}\times{\bf \eta}|^{2}),  
\label{G-G-L-C-S}
\end{eqnarray}
where ${\bf D}_{//}=(D_{x},D_{y})$.  
The Ginzburg-Landau Lagrangian Eq. (\ref{G-G-L-C-S}) is an  
extended version of the Chern-Simons-Higgs model, so to say, 
a non-relativistic Maxwell-Chern-Simons-vector Higgs model.

\section{Vortex with Fractional Charge and Fractional Angular Momentum 
in Chiral $p$-Wave Superconductor}

In this section, we study charge and angular momentum of a vortex 
in chiral $p$-wave superconductor.  
As we mentioned before, charge and angular momentum take fractional values.  
From these facts the vortex in the present case could be anyon.\cite{Wilczek}

Suppose the magnetic field is directed to the z-axis. 
We consider static and cylinder-symmetric solution around the $z$ axis  
and we neglect time and $z$-dependence. Eqs. of motion in 
${\bf \nabla} \cdot {\bf A}=0$ gauge are written as 
\begin{eqnarray}
{\bf \nabla}\cdot {\bf E}  
&=& - 8 e^{2} |{\bf \eta}|^{2} A_{0} + 
n_{p} \sigma_{xy} {\rm B}, 
\nonumber\\
({\bf \nabla} \times {\bf B})_{i}    
&=&- 8 e^{2} c_{s}^{2} |{\bf \eta}|^{2} A_{i} 
+2i e c_{s}^{2} ({\bf \eta}^{*} \partial_{i} {\bf \eta} -
{\bf \eta} \partial_{i} {\bf \eta}^{*})   
\nonumber\\
&&+2 i e c_{s}^{2} \gamma  
\epsilon_{ij3} \partial_{j} ({\bf \eta}^{*} \times {\bf \eta}) 
+ n_{p} \sigma_{xy} \epsilon_{ij3} \partial_{j} A_{0}, 
\nonumber\\  
&& ~~~~~~~~~~~~~~~~~~~~~~~~~~~~~~~~~~~~~~~~ (i=1,2) 
\nonumber\\
c_{s}^{2} \nabla^{2}\eta_{k}  
&=&-4 e^{2} ( A_{0}^{2} - c_{s}^{2} A_{//}^{2}) \eta_{k}  
-4 i e c_{s}^{2} {\bf A}_{//} \cdot {\bf \partial}_{//} \eta_{k} 
\nonumber\\
&&+ 4 i e c_{s}^{2} \gamma \epsilon_{kl} \eta_{l} {\rm B} 
\nonumber\\
&&+\frac{a}{2}(|{\bf \eta}|^{2} - |\eta_{0}|^{2})\eta_{k} 
\label{G-L-Eq}\\
&&+ 2 b \left\{|{\bf \eta}|^{2}\delta_{kl} 
- ({\bf \eta}^{*}\times{\bf \eta}) \epsilon_{kl}\right\}\eta_{l},  
\nonumber
\end{eqnarray}
where ${\rm E}^{i}=F^{i0}$ and ${\rm B}=-F^{12}$. 
We see that there are two typical length scales in these equations. 
One of them is $\lambda=(8 e^{2} c_{s}^{2} |\eta_{0}|^{2})^{-1/2}$ for 
gauge fields and another is $\xi=c_{s}(\sqrt{a}|\eta_{0}|)^{-1}$ for 
${\bf \eta}$. 
$\lambda$ coincides with so-called London penetration depth of 
the magnetic field in the case $\sigma_{xy}=0$. 
We call $\lambda$ also ``penetration depth'' in this paper. 
$\xi$ is the coherent length. The Ginzburg-Landau parameter 
$\kappa$ is defined as $\kappa=\lambda/\xi$. 
It distinguishes type I superconductor ($\kappa<\frac{1}{\sqrt{2}}$) 
and type II superconductor ($\frac{1}{\sqrt{2}}<\kappa$). 
Type II superconductor has a stable vortex solution but type I does not. 
The chiral $p$-wave superconductor $Sr_{2}RuO_{4}$ has  
values $\lambda \simeq 1800 \AA$ and $\kappa=1.2$ for the magnetic 
field directed to the c axis of the crystal, which corresponds to z axis 
in this paper.\cite{Yoshida} Therefore, $Sr_{2}RuO_{4}$ is 
a type II superconductor and it has a stable vortex.

We use cylindrical coordinate and 
consider an Ansatz written as, 
\begin{equation}
\left\{ \begin{array}{ccl}
{\bf \eta}&=&
\frac{|\eta_{0}|}{\sqrt{2}}(\rho_{1}(r),i\rho_{2}(r)) e^{i n \theta},\\
A_{0}&=&\frac{1}{2e\lambda}a_{0}(r),\\ 
A_{\theta}&=&\frac{1}{2er}(n-a_{\theta}(r)),
\end{array}\right. 
\label{Ansatz}
\end{equation}
which shows a situation that a vortex with vorticity $n$ is located  
at the origin.  For the 
total energy of the system to be finite, boundary condition at 
infinity is written as 
\begin{equation}
\left\{ \begin{array}{ccl}
\rho_{i}(\infty)&=&1~;~i=1,2,\\
a_{0}(\infty)&=&0, \\
a_{\theta}(\infty)&=&0.  
\end{array} \right. \label{bc-infty}
\end{equation}
Boundary conditions at the origin 
\begin{equation}
\left\{ \begin{array}{ccc}
\rho_{i}(0)&=&0,\\
a_{0}(0)&=&{\rm const.}, \\
a_{\theta}(0)&=&n, 
\end{array} \right. \label{bc-orig}
\end{equation}
are required to exclude singularities of the fields.

First, we consider the asymptotic behavior of Ansatz Eq.(\ref{Ansatz}). 
As we mentioned before, we see from Eq.(\ref{G-L-Eq}) that a typical 
length scale of ${\bf \eta}$ is $\xi$.  
Hence we regard $\rho_{i} \simeq 1$ when $r$ is sufficiently larger 
than $\xi$ and behavior of $a_{0}$ and $a_{\theta}$ are 
written as\cite{Goryo-Ishikawa}  
\begin{eqnarray}
a_{0}(r) &\simeq& 
C_{+} {\rm K}_{0}(\frac{r}{c_{s} \lambda})  + 
C_{-} n_{p} \sigma_{xy} \lambda c_{s}^{2} {\rm K}_{0}(\frac{r}{\lambda}) , 
\nonumber\\
a_{\theta}(r) &\simeq&  - 
\frac{r}{\lambda}   
\left\{ C_{+} c_{s} n_{p} \sigma_{xy} \lambda 
{\rm K}_{1}(\frac{r}{c_{s} \lambda})  
- C_{-} {\rm K}_{1}(\frac{r}{\lambda}) \right\},  
\label{r->infty}
\end{eqnarray} 
where K$_{l}(r)$ is the $l$-th order modified Bessel function and 
$C_{+}$ and $C_{-}$ are numerical factors.   
The power behavior of the fields near the origin is written as 
\begin{eqnarray}
a_{0} &\sim& \alpha_{0} + 
\frac{n_{p} \sigma_{xy} \lambda}{4} \alpha_{\theta}^{(2)} (r/\lambda)^{2} + 
{\cal{O}}((r/\lambda)^{4}), 
\nonumber\\
a_{\theta} &\sim& n + \alpha_{\theta}^{(2)} (r/\lambda)^{2} + 
{\cal{O}}((r/\lambda)^{4}), 
\nonumber\\
\rho_{i} &\sim& 
\rho_{i}^{(n)} (r/\lambda)^{|n|} + 
{\cal{O}}((r/\lambda)^{|n|+2}) \label{r->0},    
\end{eqnarray} 
where $\alpha_{0},\alpha_{\theta}^{(2)},\rho_{i}^{(n)}$ 
are numerical constants. 

By using asymptotic behavior of the fields, 
we calculate the charge and the angular momentum of the vortex.  
In the rest of this section, we consider 
pure 2-dimensional space to avoid complexity in the arguments. 
It corresponds to putting $n_{p}=1$ in our calculation. 

The vortex charge is obtained by integrating charge density of the matter,  
such as  
\begin{eqnarray}
Q&=&\int d^{2} x j_{0}^{Matter}
=\int d^{2}x (- 8 e^{2} A_{0} |{\bf \eta}|^{2}) 
\nonumber\\ 
&=&\int d^{2}x \left\{{\bf \nabla} \cdot {\bf E} - 
\sigma_{xy} {\rm B} \right\} 
\nonumber\\
&=&- 2 \pi \sigma_{xy} (r A_{\theta}) |_{r=0}^{r=\infty}
\nonumber\\ 
&=&- \sigma_{xy} \frac{2 n \pi}{2 e} = - \frac{n e}{4}, 
\label{fractional-charge}\end{eqnarray}
by using Eqs. (\ref{G-L-Eq}), (\ref{Ansatz}), (\ref{r->infty}) 
and (\ref{r->0}).
We can see the flux $\frac{2 n \pi}{2 e}$ is attached to the charge,  
{\it i.e.} the vortex is an object such as flux-charge composite.
Volovik has mentioned the vortex charge caused by the Chern-Simons term 
in superfluid $^{3}$He-A film. His result has half value comparing with 
our result Eq. (\ref{fractional-charge}) because the factor 2 was missing 
when the vortex charge was derived from the Chern-Simons term.
\cite{volovik-charge}  

It was argued by many authors that the change in the density of electrons 
in the vortex core due to the spatial dependence of the orderparameter 
also gives the charge of the vortex\cite{charging} 
and we call it ``the regular charge''. 
Recently, Volovik pointed out that the occupation of 
the zero-energy bound states of electrons, which exist in 
the vortex core of the chiral p-wave superconductor,\cite{Kopnin-Salomaa} 
could contributes the charge of the vortex.
\cite{Volovik-0-mode,Jackiw-Rebbi}  
We call it ``zero-mode charge''.  
It is important how to distinguish these two charges and 
the fractional charge Eq. (\ref{fractional-charge}) comes from 
the Chern-Simons term.  
It can be done, even in the case that the regular charge or 
the zero-mode charge are of order $e$, by comparing the vortex 
charges of vorticities $n=1$ and $n=-1$ because the fractional 
charge Eq. (\ref{fractional-charge}) 
changes its sign but other two charges do not depend on 
the sign of vorticity.  
 
The definition of the vortex angular momentum is  
\begin{equation} 
J=\int d^{2}x {\bf r} \times {\bf {\cal {P}}}. 
\label{angular momentum}   
\end{equation}  
${\cal {P}}$ is the momentum density, which is the 
generator of the translation. It is defined to be gauge invariant  
such as 
\begin{eqnarray}   
{\cal {P}}_{i} &=& 
\frac{\partial{\cal{L}}}{\partial (\partial_{0} {\bf \eta})} D_{i}{\bf \eta} + 
(h.c.) + 
\frac{\partial{\cal{L}}}{\partial (\partial_{0} A_{j})} F_{ij} 
\nonumber\\  
&=& (D_{0} {\bf \eta})^{*} D_{i} {\bf \eta} + D_{0}{\bf \eta} 
(D_{i}{\bf \eta})^{*} 
+ \epsilon_{ij} {\rm E}_{j} {\rm B}.  \label{momentum}
\end{eqnarray}  
By using Eqs.(\ref{G-L-Eq}), (\ref{Ansatz}), (\ref{r->infty}) 
and (\ref{r->0}), 
angular momentum Eq.(\ref{angular momentum}) is calculated as follows;   
\begin{eqnarray} 
J&=&\int d^{2}x r p_{\theta} \nonumber\\
&=& 
\int d^{2}x r \left\{ 8 e^{2} A_{0} 
(A_{\theta} - \frac{n}{2 e r}) |{\bf \eta}|^{2} - {\rm E}_{r} {\rm B} \right\} 
\nonumber\\  
&=& - 2 \pi \frac{\sigma_{xy}}{4 e^{2}} 
\int_{0}^{\infty} dr \frac{1}{2} \partial_{r} 
\left\{n - a_{\theta}(r)\right\}^{2}      
\nonumber\\
&=&-\frac{n^{2}}{16}.
\label{fractional value}
\end{eqnarray}  

We see the vortex in our system Eq.(\ref{G-G-L-C-S}) has fractional 
charge and fractional angular momentum same as the vortex  
in Chern-Simons-Higgs model.\cite{A-H-C-S} 
Since the charge and the angular momentum are proportional to $\sigma_{xy}$,  
we see that the Chern-Simons term creates 
these non-zero fractional values. 
These values are topological and depends only on asymptotic behavior of 
the gauge fields, although 
there is Maxwell term $-\frac{1}{4}F^{\mu\nu}F_{\mu\nu}$ 
and the Higgs field is vector-like. 

The results Eq. (\ref{fractional-charge}) and 
Eq. (\ref{fractional value}) would be general and 
would not depend on the form of Ansatz and phenomenological parameters. 
Actually, in unconventional superconductors 
there is another Ansatz\cite{Volovik-Vachaspati} 
which is different from Eq. (\ref{Ansatz}),  
and there are plausible phenomenological parameters for $Sr_{2}RuO_{4}$ 
studied by Agterberg\cite{Agterberg} and they are different from 
Eq. (\ref{parameters-Sig.-Ueda}). 
The results Eq. (\ref{fractional-charge}) and 
Eq. (\ref{fractional value}) would not be changed 
even if we use such a different Ansatz and phenomenological parameters.  
The reasons are written as follows; 

\begin{enumerate} 

\item The definition of charge density and angular momentum density 
do not depend on the phenomenological parameters {\it explicitly} 
and always become total derivative, since 

\begin{enumerate}

\item charge density of matter $j_{0}^{Matter}$ 
in Eq. (\ref{fractional-charge}) always satisfies the Maxwell equation 
$j_{0}^{Matter}={\bf \nabla} \cdot {\bf E} - \sigma_{xy} {\rm B}$, and

\item the momentum density ${\cal {P}}_{i}$ defined by Eq. (\ref{momentum}) 
depends on the form of time derivative terms in Lagrangian 
Eq. (\ref{G-G-L-C-S}) and these terms do not contain the phenomenological 
parameters. 

\end{enumerate}

\item Asymptotic behavior of the fields is   
investigated without the Chern-Simons term 
in Ref. \cite{Heeb-Agterberg} by using 
Ansatz written in Ref.\cite{Volovik-Vachaspati} and 
the phenomenological parameters written in Ref.\cite{Agterberg}.   
The essential property of the asymptotic behavior is 
the same with Eq. (\ref{r->infty}) and Eq. (\ref{r->0}), 
and it would not be changed even if the Chern-Simons term 
is taken into account.

\end{enumerate}

Therefore, the results Eq. (\ref{fractional-charge}) and 
Eq. (\ref{fractional value}) would be valid for realistic 
systems such as superconducting $Sr_{2}RuO_{4}$. 
Evidence that the vortex we consider is a candidate of anyon 
can be seen in the fact that vortex has fractional 
angular momentum.\cite{Wilczek} 
It is expected that transmutation of the statistics occurs.  
The further discussion is needed 
to see the fractional statistics of these vortices experimentally.

%Next, we study the statistics of the vortices. 
%We naively define one vortex wave function $\varphi({\bf r})$, 
%where ${\bf r}$ shows the vortex center.  
%Since the vortex has angular momentum $J$, we write down 
%azymuthal dependence as $\varphi({\bf r}) \sim e^{i J \arg({\bf r})}$.  
%We consider two identical vortices with their coordinates  
%${\bf r}_{1}$ and ${\bf r}_{2}$. 
%Each of them has a vorticity $n$. 
%Suppose there is no interaction between them, which may be reasonable   
%when the distance of these two vortices is much larger than the penetration 
%depth $\lambda$. 
%The interchange of these two vortices 
%${\bf r}_{1} \leftrightarrow {\bf r}_{2}$ is realized by rotate them  
%around the middle point 
%$\frac{{\bf r}_{1} + {\bf r}_{2}}{2}$ 
%with rotation angle $\pi$. 
%When we put the middle point on the origin,  
%the azymuthal dependence of 
%the two-vortex wave function $\Phi$ is written as 
%\begin{equation}
%\Phi({\bf r}_{1},{\bf r}_{2}) \sim e^{2 i J \arg({\bf r})},  
%\end{equation} 
%where ${\bf r}={\bf r}_{1} - {\bf r}_{2}$. 
%After rotating them, a phase factor arises 
%\begin{eqnarray}
%\Phi({\bf r}_{2},{\bf r}_{1})&=&e^{2 i \pi J} \Phi({\bf r}_{1},{\bf r}_{2})
%\nonumber\\
%&=&e^{i \frac{\pi n^{2}}{8}} \Phi({\bf r}_{1},{\bf r}_{2}). 
%\end{eqnarray}
%Evidence that the vortex we consider is a candidate of anyon 
%can be seen in above calculation\cite{Wilczek}. 
%It is expected that transmulation of the statistics occurs 
%and the further discussion is needed 
%how to see the fractional statistics of these vortices experimentally. 

\section{Electromagnetic Property of Vortex 
with Fractional Quantum Numbers} 

In this section, we show numerical solution of 
Eq. (\ref{G-L-Eq}) by using Ansatz Eq. (\ref{Ansatz}) and discuss  
electromagnetic property of the vortex. We 
consider layered system again. In this case, 
$n_{p} \simeq 10^{-1}\AA^{-1}$. 
For simplicity, we assume  
$\gamma$ in Eq. (\ref{G-G-L-C-S}) is zero.   
In this case, there is an exchange symmetry such as  
$\rho_{1} \leftrightarrow \rho_{2}$ in the equation which is obtained by 
substituting Eq. (\ref{Ansatz}) into Eq. (\ref{G-L-Eq}). 
We introduce a real field $\phi$ written as 
\begin{equation} 
\phi(r)=\rho_{1}(r)=\rho_{2}(r).    
\end{equation}
The magnetic field, the electric field and $\phi$ 
with parameters $\kappa=1.2$ and  
$c_{s} \simeq (\frac{{v_{\rm F}}^{2}}{2 \pi})^{1/2} \simeq 10^{-4}$ 
in the case $n=1$ are shown in FIG. 1.
%%%%%%%%%%%%%%%%%%%%%%%%%%%%%%%%%%%%%%%%%%%%%%%%%%%%%%%%%%%%%%%%
%From asymptotic behavior of $a_{0}$ and $a_{\theta}$ in Eq. (\ref{r->infty}), 
%we see there is a relation $A_{0}/ A_{\theta} \simeq c_{s}^{2}$. 
%As we mentioned before in section II, 
%$c_{s} \simeq (\frac{{v_{\rm F}}^{2}}{2 \pi})^{1/2} \simeq 10^{-4}$.  
%Therefore, for sufficiently large $r$, Eq. (\ref{G-L-Eq}) reduces to 
%\begin{eqnarray}
%\frac{1}{c_{s}^{2} \lambda^{2}} \phi^{2} A_{0}
%&\simeq&n_{p} \sigma_{xy} \frac{1}{r}\frac{d}{d r}(r A_{\theta}), 
%\label{A0}\\
%-\frac{d}{d r} \frac{1}{r}(\frac{d}{d r} r A_{\theta}) 
%&\simeq&
%- \frac{1}{\lambda^{2}} \phi^{2} (A_{\theta} - \frac{n}{2 e r}), 
%\label{At}\\  
%-\frac{1}{r}\frac{d}{d r} (r \frac{d}{d r} \phi) 
%&\simeq&4 e^{2} 
%(A_{\theta} - \frac{n}{2 e r})^{2} \phi -\xi^{2}(\phi^{2} - 1)\phi. 
%\label{phi}
%\end{eqnarray}
%Eq. (\ref{At}) and Eq. (\ref{phi}) coincide with the usual 
%vortex equation and we solve them numerically with parameter 
%$\kappa = \frac{\lambda}{\xi}=1.2$ 
%and we find $A_{0}(r)$ by substituting these solutions into Eq. (\ref{A0}).
%%%%%%%%%%%%%%%%%%%%%%%%%%%%%%%%%%%%%%%%%%%%%%%%%%%%%%%%%%%%%%%%%
Contrary to the usual vortex ({\it i.e.} the vortex in the case 
$\sigma_{xy}=0$), we found that an electric field in the radial direction 
is induced in the region 
$r < \lambda$ and its magnitude is about $1$ Volt/meter.

Let $\Delta E$ stands for energy difference 
between our vortex $E(\sigma_{xy}=\frac{e^{2}}{4 \pi})$ and 
the usual one $E(\sigma_{xy}=0)$. There is a relation between them written as  
\begin{eqnarray}
\frac{\Delta E}{E(\sigma_{xy}=0)}
&=&\frac{E(\sigma_{xy}=\frac{e^{2}}{4 \pi})}{E(\sigma_{xy}=0)} - 1
\nonumber\\
&\sim& 10^{-8}.  
\end{eqnarray}
Therefore, energy increase caused by the existence of the Chern-Simons term 
is extremely small and the vortex 
with fractional quantum numbers could exist in realistic systems.   
The magnitude of the vortex energy  
and also that of the electric field depend on the form of Ansatz and 
phenomenological parameters in Eq. (\ref{G-G-L-C-S-0}), however, 
these calculation would have validity as an order estimation.  
It is interesting if some phenomena caused by the induced 
electric field are detected by experiments.

\section{Conclusion} 

We have discussed a vortex in chiral $p$-wave superconductor such as 
$Sr_{2}RuO_{4}$, where $U(1)$, P- and T-symmetry are broken simultaneously.  
We have investigated the vortex 
based on the Ginzburg-Landau Lagrangian.  
The Ginzburg-Landau Lagrangian of the system contains 
the Chern-Simons term,\cite{Goryo-Ishikawa} 
and it corresponds to an extended model of the Chern-Simons-Higgs model. 
We have found that the vortex in chiral $p$-wave superconductor 
has fractional charge $-\frac{n e}{4}$ and 
fractional angular momentum $-\frac{n^{2}}{16}$ with integrally 
quantized magnetic flux $\frac{2 n \pi}{2 e}$ . These values are 
topologically stable and do not depend on the form of Ansatz and 
the phenomenological parameters in the Ginzburg-Landau Lagrangian. 
Following the discussion in Ref.\cite{Wilczek}, the vortex 
could obey the fractional statistics. 
We have also investigated the electromagnetic property of the 
vortex. We have found that the electric field is induced by 
the Chern-Simons term near the vortex core. 
Energy increase caused by the existence of the Chern-Simons term 
is so small that the vortex with fractional quantum numbers could exist 
in realistic systems. It is interesting if these exotic feature is 
observed experimentally. 

Our discussion is valid for other chiral superconductors, 
such as ``anisotropic chiral $p$-wave'' superconductor in which 
the symmetry of the wave function of Cooper pair is 
$(\sin p_{x} + i \sin p_{y})$-wave \cite{Miyake-Narikiyo} 
and $d_{x^{2}-y^{2}}+id_{xy}$-wave superconductor,\cite{Laughlin2} 
since it has been shown that the Chern-Simons term 
is also induced in the Ginzburg-Landau 
Lagrangian of these superconductors. 
\cite{Goryo-Ishikawa}

\acknowledgements

The author is grateful to Professor K. Ishikawa and Dr. N. Maeda 
for their suggestion to consider this subject, for helpful 
discussions, and for critical reading of the manuscript. 
The author also thanks to Professors 
Y. Maeno, M. Matsumoto, M. Sigrist, G.E. Volovik and 
Dr. Y. Okuno for useful discussions. 
This work was partially supported by the special Grant-in-Aid for 
Promotion of Education and Science in Hokkaido University provided by 
the Ministry of Education, Science, Sports, and Culture, the Grant-in-Aid 
for Scientific Research on Priority area(Physics of CP violation)
(Grant No. 10140201), and the Grant-in-Aid for International Science Research 
(Joint Research 10044043) from the 
Ministry of Education, Science, Sports and Culture, Japan.

\end{multicols}

\newpage

\begin{center}
FIGURE CAPTION
\end{center}

FIG. 1. $r$ dependence of (a) the 
orderpapameter $\phi$, (b) the magnetic field directed to the $z$-axis, 
and (c) the radial electric field.  

%\newpage

\begin{center}
FIGURES
\end{center}

%%%%%%%%%%%%%%%%%%%%%%%%%%%%%%%%%%%%%%%%%%%%%%%%%%%%%%%%%%%%%%%%%%%%%%%%
\vspace{-2cm}
\begin{figure}
\centerline{
\epsfysize=20cm\epsffile{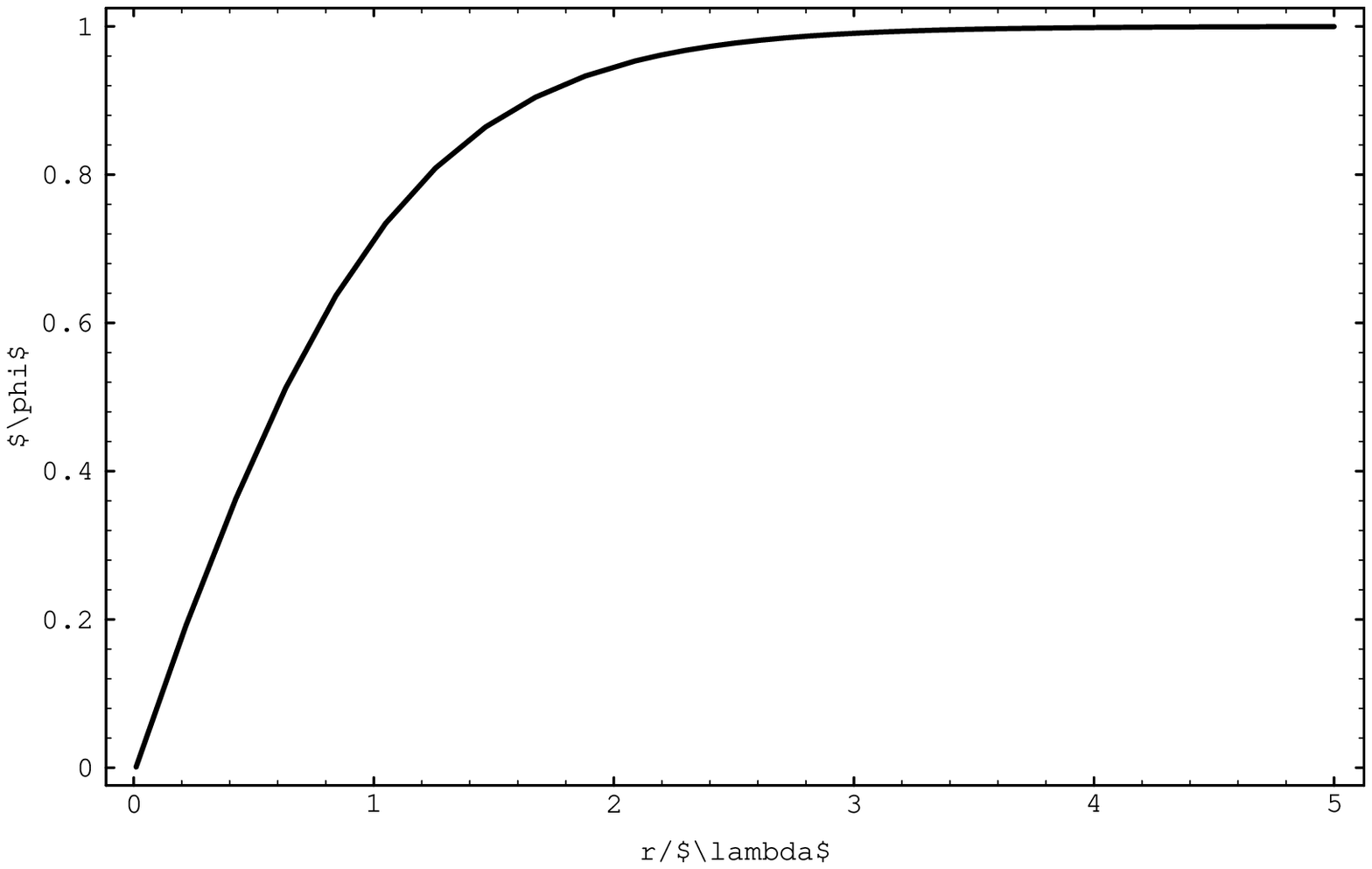}}
\vspace{-2cm}
FIG.~1(a). 

\end{figure}
%%%%%%%%%%%%%%%%%%%%%%%%%%%%%%%%%%%%%%%%%%%%%%%%%%%%%%%%%%%%%%%%%%%%%%%%%
%%%%%%%%%%%%%%%%%%%%%%%%%%%%%%%%%%%%%%%%%%%%%%%%%%%%%%%%%%%%%%%%%%%%%%%%
\vspace{-2cm}
\begin{figure}
\centerline{
\epsfysize=20cm\epsffile{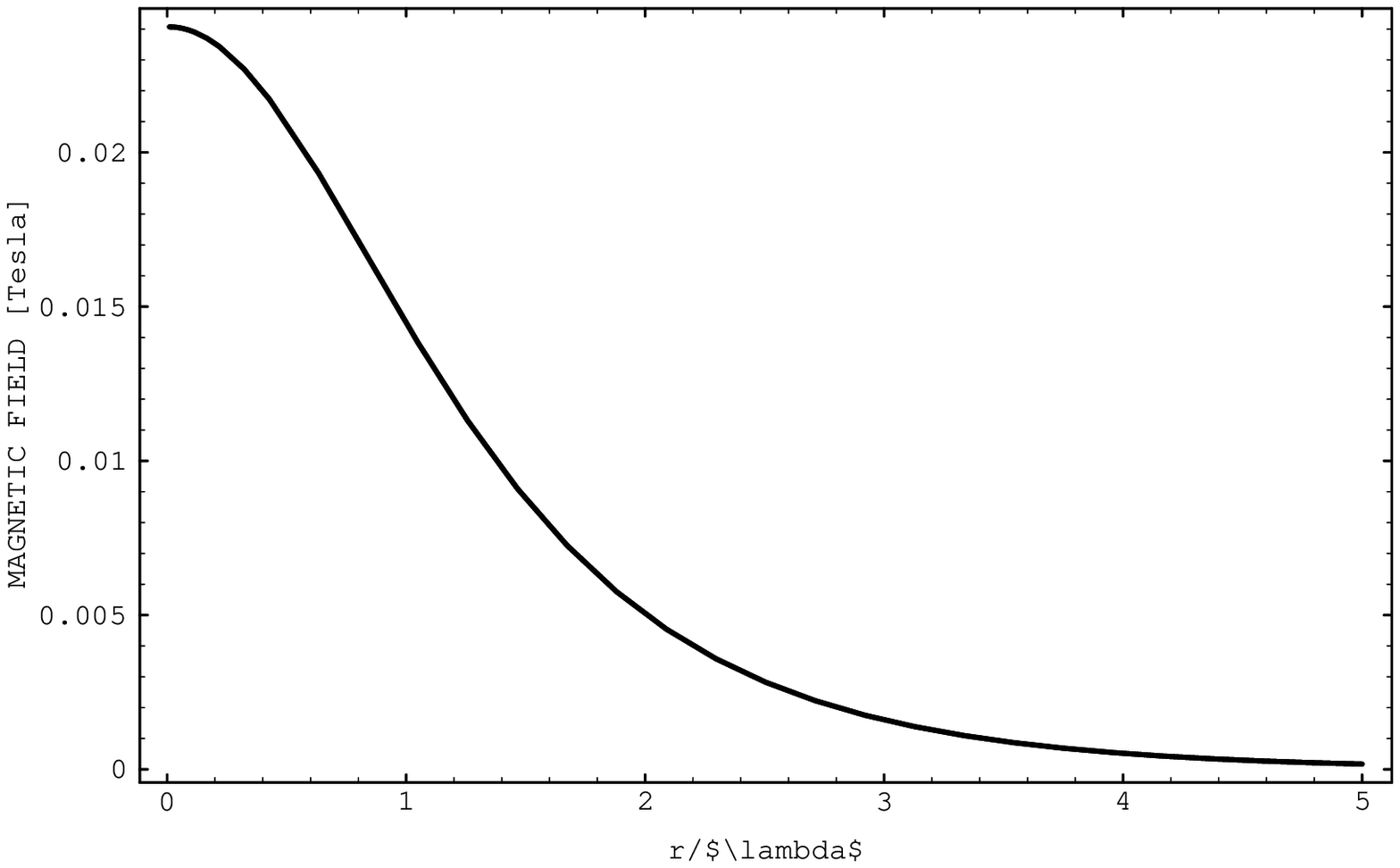}}
\vspace{-2cm}
FIG~1(b). 

\end{figure}
%%%%%%%%%%%%%%%%%%%%%%%%%%%%%%%%%%%%%%%%%%%%%%%%%%%%%%%%%%%%%%%%%%%%%%%%%
%%%%%%%%%%%%%%%%%%%%%%%%%%%%%%%%%%%%%%%%%%%%%%%%%%%%%%%%%%%%%%%%%%%%%%%%
\vspace{-2cm}
\begin{figure}
\centerline{
\epsfysize=20cm\epsffile{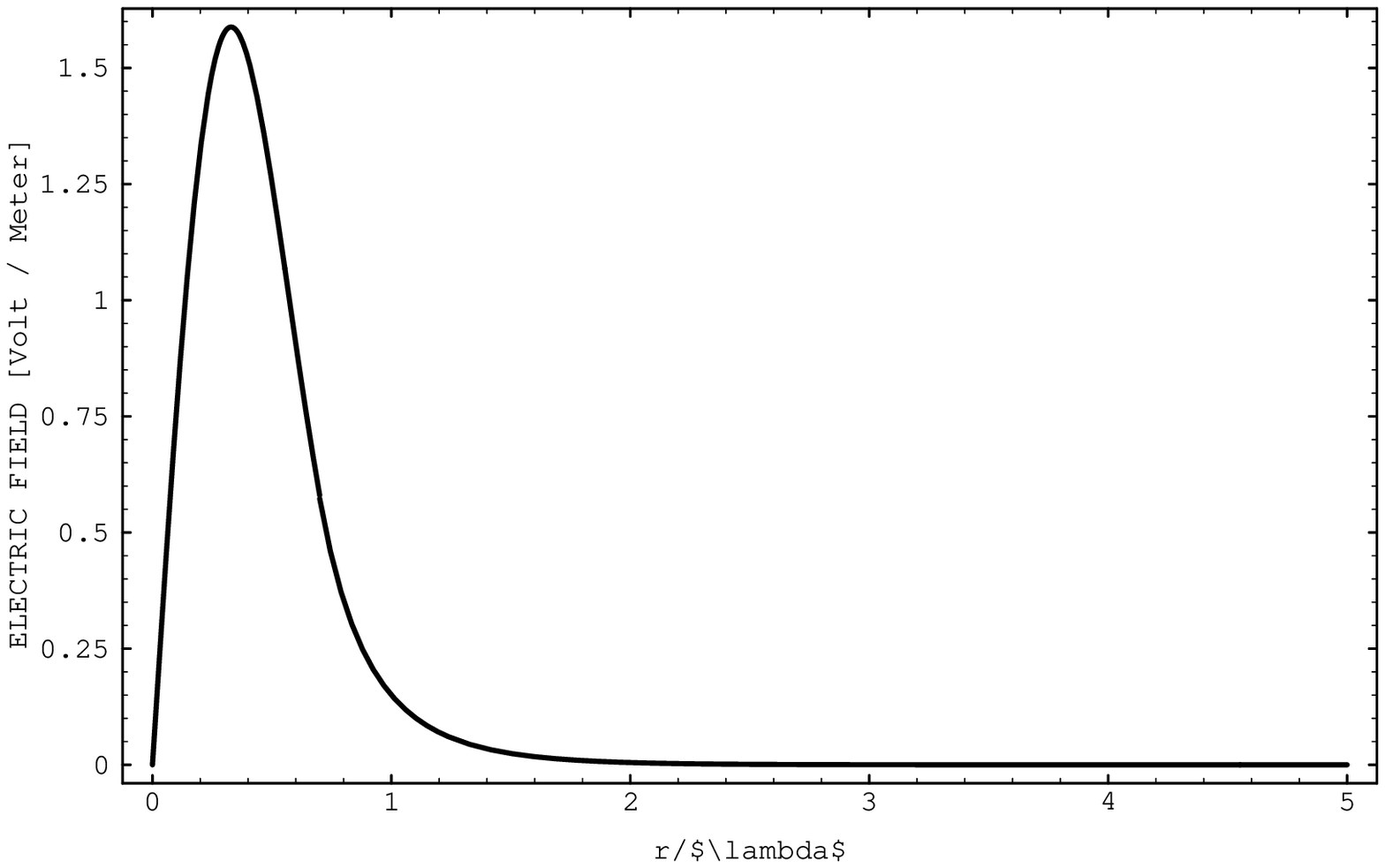}}
\vspace{-2cm}
FIG.~1(c). 

\end{figure}
%%%%%%%%%%%%%%%%%%%%%%%%%%%%%%%%%%%%%%%%%%%%%%%%%%%%%%%%%%%%%%%%%%%%%%%%%

\end{document}